\documentclass[preprint,12pt]{elsarticle}
\usepackage{amsmath}
\usepackage{amsfonts}
\usepackage{amssymb}
\usepackage{graphicx}
\usepackage{rotating}
\usepackage{amsbsy}

\journal{Medical Engineering and Physics}

\begin{document}

\begin{frontmatter}



\title{Lung Function Measurement with Multiple-Breath-Helium Washout System}


\author[a,b]{J.-Y.~Wang}
\ead{jauyi.wang@kcl.ac.uk}
\author[a]{J.R.~Owers-Bradley}
\ead{john.owers-bradley@nottingham.ac.uk}
\author[a]{M.E.~Suddards}
\ead{matt.suddards@nottingham.ac.uk}
\author[a]{C.E.~Mellor}
\ead{chris.mellor@nottingham.ac.uk}
\address[a]{School of Physicis and Astronomy,University of Nottingham, University Park, Nottingham, UK, NG7 2RD\fnref{fn1}}
\address[b]{Engineering Division, King's College London, Strand, London, UK, WC2R 2LS\corref{cor1}}

\begin{abstract}
Multiple-breath-washout (MBW) measurements are regarded as a sensitive technique which can reflect the ventilation inhomogeneity of
respiratory airways. Typically nitrogen is used as the tracer gas and is washed out by pure oxygen in multi-breath-nitrogen (MBNW) washout tests. In this work, instead of using nitrogen, $^4$He is used as the tracer gas and a multiple-helium-breath-washout (MBHW) system has been developed for the lung function study. A commercial quartz tuning fork with a resonance frequency of 32768 Hz has been used for detecting the change of the respiratory gas density. The resonance frequency of the tuning fork decreases linearly with increasing density of the surrounding gas. Knowing the CO$_2$ concentration from the infrared carbon dioxide detector, the helium concentration can be determined. Results from 12 volunteers (3 mild asthmatics, 2 smokers, 1 with asthma history, 1 with COPD history, 5 normal) have shown that mild asthmatics have higher ventilation inhomogeneity in either conducting or acinar airways (or both). A feature has been found in single breath washout curve from 4 smokers with different length of smoking history which may indicate the early stage of respiratory ventilation inhomogeneity.
\end{abstract}


\end{frontmatter}


\section{Introduction}
\label{Introduction}

\subsection{Physiological Background}

Multiple-breath washout (MBW) is regarded as a sensitive technique to study
the ventilation inhomogeneity in conducting and acinar airways ~\cite{Engel_book,SylviaCOPD,newth1997,Sylvia2007,Sylvia2004,Aurora2005,Monika2006}. In the majority of MBW measurements, subjects breathe in pure oxygen gas, replacing the air and washing nitrogen gas (the tracer gas) out of lungs breath by breath. The concentration of the tracer gas along with the respiratory tidal volume is carefully monitored as shown in figure 1. By plotting out the concentration of the tracer gas versus the expired tidal volumes, each single breath can be divided into three phases as shown in the inset of figure 1 uppergraph. Phase I corresponds to the pure inspiratory gas and there is no tracer gas. Phase II, with a sharp increase of gas concentration, is the transition between the inspiration and the gas remaining in the lung ~\cite{Fowler1948}. Phase III corresponds to the alveolar plateau with a positive slope ~\cite[pp.\,296]{Engel_book}.

\begin{figure}[tbp]
\centering 
\includegraphics[width = \textwidth]{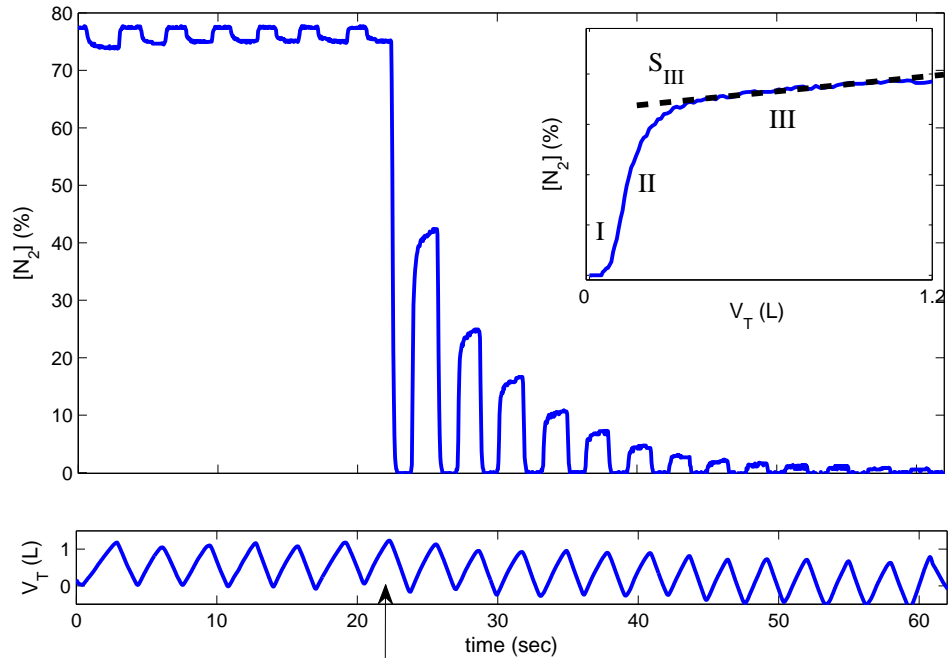} 
\caption[Multiple-breath nitrogen washout curve]
{The typical nitrogen washout curve measured in University of Leicester from a volunteer with a commercial nitrogen washout system (Ultima PF, Medical Graphics Cooperation, USA). The upper graph shows the concentration of nitrogen and the lower graph shows the respiratory tidal volumes during the washout process. The arrow indicates the start of inspiring pure oxygen. The inset in the uppergraph shows a single breath curve where the nitrogen concentration is plotted against the tidal volume. I, II and III represent phases I, II and III. S$_{III}$ represents the phase III slope which is fitted by a straight dashed line.}
\label{Multiple-breath nitrogen washout curve}
\end{figure}

The analysis of MBW focusses on phase III. The mechanism which causes the phase III slope to be slightly positive is inhomogeneous ventilation due to the asymmetrical structure of alveolar airways and the uneven distribution of alveoli. The inspired gas with lower concentration is mixed with the gas remaining in alveolar sacs and is distributed unevenly to different units of the lungs ~\cite{Fowler1949,Fowler1951,Paiva_Engel1984,Paiva_Engel1987,Paiva_eq}. After an inspiration, the gas mixture in those respiratory units with less ventilation, (i.e., more difficult for the inspired gas to diffuse into,) has a higher concentration of tracer gas and is flushed out less easily. The gas mixture in the better-ventilated units relatively has a lower concentration of tracer gas since the gas is diluted well and contributes to the early part of the alveolar expiration. Less-ventilated regions contribute to the later part of the phase III which causes the positive phase III slope. The washout results from individuals with lung diseases have steeper phase III slopes compared to normal subjects because of the greater inhomogeneity of their respiratory ventilation ~\cite{Fowler1949,Paiva1982,Paiva1984,Crawford1987,Crawford1985}. Other minor effects such as body posture or inspiration and expiration flow rates can also increase the phase III slopes ~\cite{Paiva1984,Paiva1995}.

\subsection{Physiological Analysis}

\subsubsection{Lung Clearance and Two Compartment Model}

During washout, for unevenly ventilated lungs, the alveolar gas
concentration becomes unequal from unit to unit and the first breath in the
washout should reflect the gas flow the most from the better-ventilated units. The
degree of ventilated inhomogeneity arises progressively less from those
better-ventilated units and more from the poorly-ventilated units in the
second and subsequent breaths. This is because the alveolar gas is washed
out more quickly from the better-ventilated units than from
poorly-ventilated units.

For an ideal container with a perfect ventilation, i.e., the inspired
diluting gas mixes perfectly well with the tracer gas remaining in it, the
mean concentration of expiration is exponentially related to the breath
number in the MBW test with constant tidal volumes. The relation can be
given by ~\cite{newth1997,Paiva1975} 
\begin{equation}
c_{n}=c_{0}\left( \frac{FRC}{FRC+V_{T}}\right) ^{n}
\end{equation}%
where $n$ is the breath number, $c_{n}$ and $c_{0}$ are the tracer gas
concentration from the $n$th breath and from the lungs before washout
starts, $FRC$ is the residual capacity (or the volume of the container
before inspiration), and $V_T$ is the volume of tidal breath. Equation 1 can be written as 
\begin{equation}
log\ c_{n}=log\ c_{0}+nlog\ \frac{FRC}{FRC+V_{T}}.
\end{equation}

Hence for the ideal lung with steady respiratory tidal volumes, the logarithm of the mean concentration is
linearly related to the breath number. However, for real lungs, the uneven distribution of inspired gas is washed out asynchronously and it results in a nonlinear relationship between $logc_{n}$ and $n$. The interpretation of the washout curve from MBNW was first studied by Robertson et al. ~\cite[pp.\,312]{Engel_book}~\cite{Robertson1950}. The washout curve corresponds to the sum of two (or more) exponential functions representing two (or more) compartments independently ventilated at different rates as shown in figure 2(b). For patients with airway diseases, the respiratory ventilation inhomogeneity reflects in a more curved line ~\cite{Lutchen1990}. 

\begin{figure}[tbp]
\centering  
\includegraphics[]{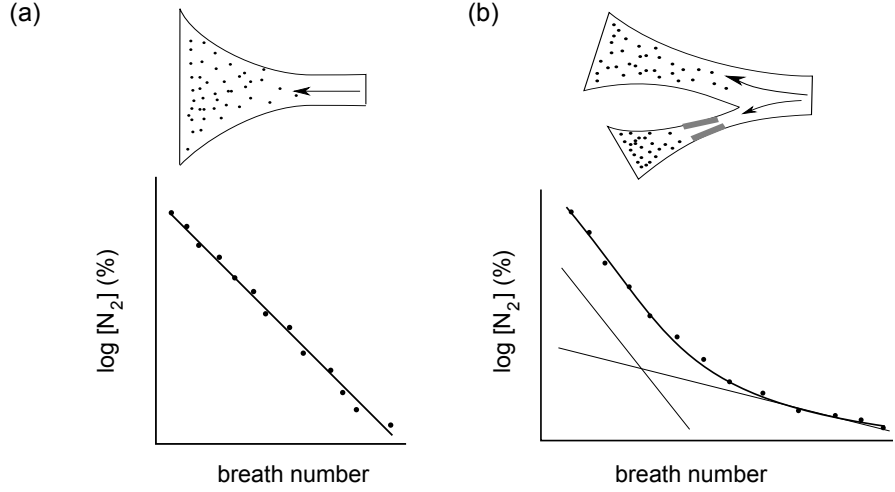}
\caption[Lung Model and Washout Curves]{(a) The perfect
lung model as a well-ventilated container with perfect gas mixing. The
multiple-breath nitrogen washout curve is a straight line. (b) Two (or more)
compartment model with two different ventilated rates. The nitrogen washout
curve is a sum of two exponential curves~\cite{Bouhuys1956}~\cite[pp.\,312]{Engel_book}.}
\label{Two Compartment Model}
\end{figure}

The lung contains units with different ventilation rates and the tidal volume also varies breath by breath. In this study, the lung is simply modelled as two compartments with different ventilation rates leading to lung clearance curves that are fitted by the sum of two exponential curves with respect to the turnover $TO$ ~\cite{Cumming1967}. The turnover is defined as the cumulative expired tidal volumes divided by functional residual capacity $FRC$, i.e., $TO_n=\sum_1^nV_T/FRC$. The decay rates and the volume ratio of the two compartments are also compared for different subjects. Equation 1 is then modified to 
\begin{equation}
c_{n}=c_{0}\left( \frac{FRC}{FRC+V_{T}}\right)^{TO}.
\end{equation}

Considering the two compartments $1$ and $2$ with the capacities $FRC_1$, $FRC_2$ and tidal volumes $V_1$, $V_2$, the lung clearance mechanism can be given by
\begin{eqnarray}
FRC &=& FRC_1+FRC_2, \\
V_T &=& V_1+V_2, \\
c(1+2,n) &=& \frac{FRC_1}{FRC_1+FRC_2}c_{1,n}+\frac{FRC_2}{FRC_1+FRC_2}c_{2,n}, \\
c_{1,n} &=& c_{a0}\left(\frac{FRC_1}{FRC_1+V_1}\right)^{TO}=c_{a0}R_1^{TO}, \\
c_{2,n} &=& c_{a0}\left(\frac{FRC_2}{FRC_2+V_2}\right)^{TO}=c_{a0}R_2^{TO},
\end{eqnarray} 
where $c_{n=0,1,2,\cdots n-1}$ are the mean concentration of gas and $V_{T,n (n=0,1,2,\cdots n-1)}$ are the tidal volumes from the initial gas in the lungs, the first, second, and $(n-1)th$ expiration, respectively. $R_1$ and $R_2$ represent the washout rates from those two compartments. The more details of the calculation is given in Appendix A.

The lung clearance index ($LCI$) has been commonly used as a measure of lung health and is defined as the total expired volume during washout divided by the $FRC$, for a fall in tracer gas concentration of a factor 40, i.e., the maximum turnover of the gas concentration above about 2\% ~\cite{Horsley2009,HorsleyCF,LCI2008}. It has been shown that the $LCI$ is more sensitive than the spirometry in cystic fibrosis (CF) adults ,and its mean value has been measured 13.1$\pm$3.8 in CF patients and a much lower mean value of 6.7$\pm$0.4 in normal subjects~\cite{Bouhuys1963,Cutillo1972,Cutillo1981,Horsley2009,HorsleyCF,LCI2008,Kraemer2005,Kraemer2008}. Other studies have also shown the increasing $LCI$ with the age ~\cite{Bouhuys1963} and a higher range of $LCI$ in emphysematous patients ~\cite{Becklake1951}. In this study, we compare the $LCI$ values from 12 volunteers; however, the $LCI$ values have not been found to differ significantly between the mild asthmatics or smokers and the normal subjects. 

\subsubsection{Two Indices $S_{cond}$ and $S_{acin}$}

Each breath of the MBW can be regarded as a single washout curve composed of
phase I to III. By analysing the phase III slopes ($S_{III}$) from each breath,
the washout results provide two indices $S_{cond}$ and $S_{acin}$ which
reflect the degree of ventilation inhomogeneities in conducting and acinar
airways, respectively. For a perfect ventilated lung, the phase III is flat
and S$_{III}$ is near zero. In real lungs, $S_{III}$ is not only positive but also the normalised value, i.e., $S_{III}$ devided by the mean tracer gas concentration, increases breath by breath. $S_{cond}$ value is taken as the increasing rate of the noramlised phase III slopes after the first few breaths. The higher $S_{cond}$ value reflects the greater ventilation inhomogeneity in conductiing airways~\cite{Crawford1985,Crawford1987,Paiva1979,Paiva1973,Sylvia1997}. $S_{acin}$ is the nomralised phase III slope from the first breath with the subtraction of $S_{cond}$, i.e.,

\begin{equation}
S_{acin}=S_{III}(1)-S_{cond}\cdot TO
\end{equation}
The higher $S_{acin}$ value the greater ventilation inhomogeneity in acinar airways. 
More details of the physiological theory on $S_{cond}$ and $S_{acin}$ can be
found in ~\cite{SylviaCOPD,Sylvia2007,Sylvia2004}.

Figure 3 shows the normallised phase III curves of helium washout measurements
from a normal and a mild asthmatic volunteer. The normalised phase III slopes ($S_{iii}$) of the asthmatic subject is apparently higher than of the normal subject as well as the increasing rate of the $S_{iii}$, i.e.,  the larger $S_{cond}$ value, which implies the higher ventilation inhomogeneity in the conductive airways.

\begin{figure}[tbp]
\centering  
\includegraphics[width = 10cm]{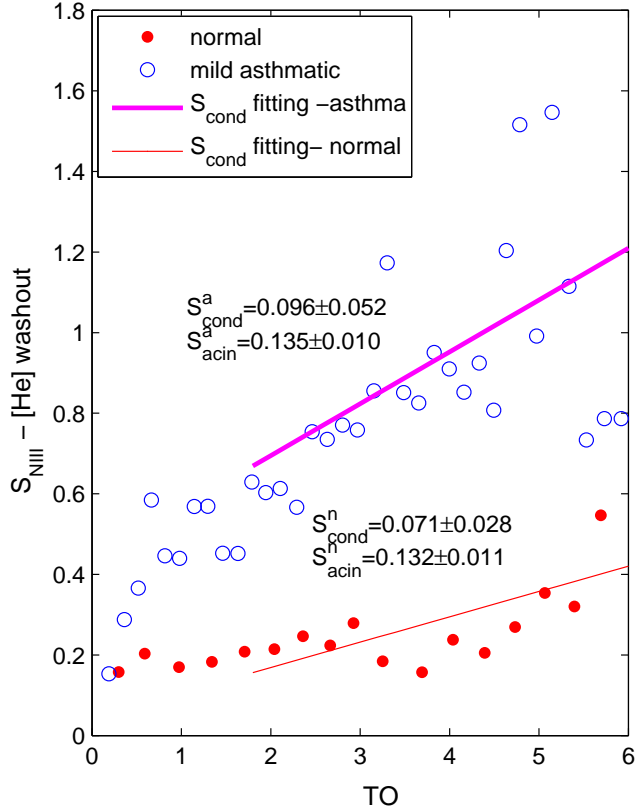}
\caption[Normalised Phase III Slope Helium Washout Curves - Normal and Asthmatic]{
The helium washout results of normalised phase III slopes versus
turnover from a normal subject and a mild asthmatic volunteer.}
\label{Normalised Phase III Slopes from Normal A Subject and A Mild Asthmatic Volunteer}
\end{figure}

\subsection{Multiple-Breath-Washout Measurements}

Nitrogen is usually chosen as the tracer (MBNW) because it is the gas we
normally breathe and has no direct influence on physiology unlike oxygen. SF$_{6}$ is also becoming more frequently used in low concentrations ~\cite{Monika2006,Horsley2009,HorsleyCF} but in this study, $^{4}$He gas is used as the tracer gas (MBHW) and is washed out by room air. Since helium not only has smaller density which is frequently mixed with oxygen to aid breathing but also has higher diffusivity than nitrogen, it is belived to be able to reach deeper into our lungs in a given time. Therefore, helium washout may be a more sensitive maneuver reflecting the ventilation inhomogeneity especially in small airways compared to nitrogen washout. The other important point is that $^{4}$He does not diffuse through the airway wall into the blood vessels but is washed out of the lungs by air.

In our MBHW system, a quartz tuning fork (QTF) with a resonance frequency
32768 Hz is used as the gas density sensor. The resonance frequency of the
QTF is linearly related to the surrounding gas density~\cite{Chu_TF,Sader1998,Sader2006}. The helium
concentration is calculated from the tuning fork signal having subtracted
the contributions to the density of all other gas components, including
carbon dioxide, oxygen, nitrogen, and water vapour. In our system, the
carbon dioxide is detected by the infrared sensor and the water is filtered
out by an ice bath. The oxygen concentration is given by the assumption of a
constant gas exchange rate from the pre-washout measurement. The helium washout curve is shown in figure 4. 

We present results on 14 volunteers to demonstrate typical signals and to illustrate the information that can be derived from the MBW. We have performed the washout measurements on three mild asthmatics, four smokers, two having mild asthma or COPD history, and six five subjects. The peak expiratory flow is also measured for each subject.

\begin{figure}[tbp]
\centering 
\includegraphics[width = \textwidth]{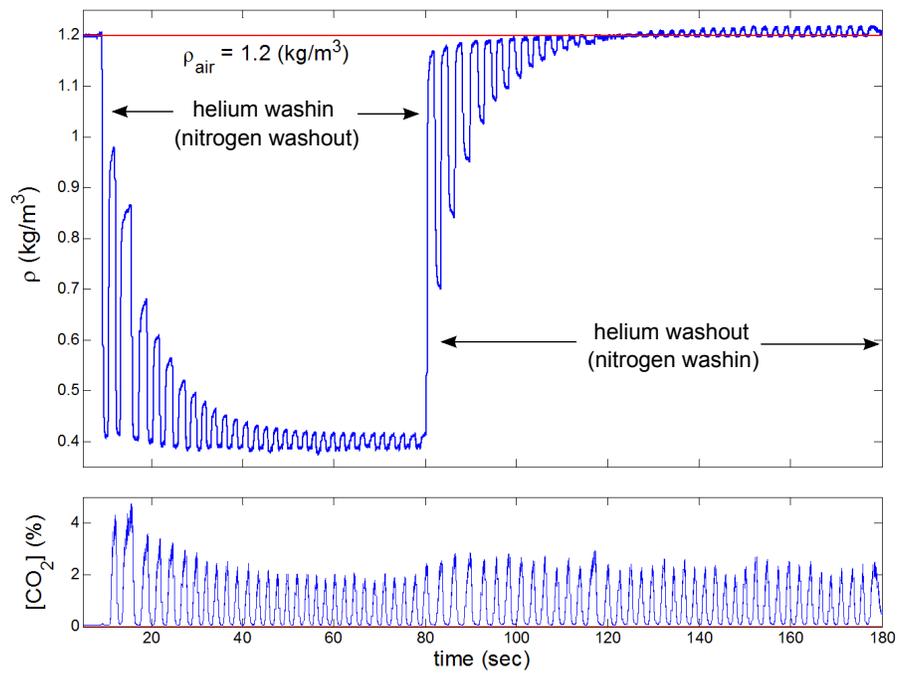}
\caption[MBHW Curve]{Multiple-breath-helium washout (MBHW) curve. It consists of two stages- washing in helium and helium washout. In helium wash-in stage, the heliox gas is washed into the lungs and washed out the room air. In the helium washout stage, the lungs are full of heliox and washed out by room air. }
\label{MBHW Curve}
\end{figure}

\section{Materials and Methods}

\subsection{Principle of Measurement}
The multiple-breath washout system is designed to analyse respiratory gas
concentrations and measure the gas volumes. Measuring the helium gas
concentration instantly, accurately and at relatively low cost is a
challenge. In this study, the MBHW measurement is based on two physiological
assumptions:
\begin{itemize}
\item Helium gas diffusing through the blood vessels is considered to be
zero.
\item The gas exchanging rate(d[O$_2$]/d[CO$_2$]), i.e., oxygen consumption
versus carbon dioxide production, remains steady for different breaths and
the existence of helium gas will not influence the gas exchange rate.
\end{itemize}

The second assumption is based on our measured results of two linear
relations between oxygen concentration and carbon dioxide concentration in
each tidal expired breath before inhaling heliox gas. Also, the concentration of carbon dioxide
remains indenpendent of the existence of helium gas which has been
observed from our measurements. 

The MBHW measurements can be divided into wash-in and washout parts. During the
first part, wash-in, heliox gas (21\% oxygen, 79\% helium) is inhaled into
the lungs and the air is washed out of the lungs breath by breath. During
the second part, washout, heliox gas is washed out of lungs by room air.
Before the wash-in, a few normal tidal breaths are taken to measure the gas
exchanging rate.

\subsection{System Design}
The MBHW system monitors the helium tracer gas concentration and the
respiratory flow rate breath by breath. Thus the system can be devided into
a gas flow measuring sub-system and the gas analysing sub-system as shown in
figure 5. The gas flow measuring sub-system contains two flow
meters to measure the inspired and expired gas flow with a pressure sensor
to monitor the direction of the flow at the mouth.

A small amount of respiratory air is pumped through the gas analysing
sub-system, which mainly consists of a QTF gas density detector, and an
infrared (IR) CO$_{2}$ detector. Since the expired gas mixture from our
lungs is saturated with water vapour $[H_{2}O]$ which interferes with the
QTF and IR detector signal, the water vapour is removed by passing through
an ice bath and reduced to around 0.6 \% of the total gas concentration, according to the Goff-Gratch (1946) equation. No moisture could be detected with a commercial moisture meter (The Shaw Automatic Dewpoint Meter, Shaw, UK). The QTF resonance frequency ($f_{c}$) gives a big response to helium gas (with about 4 Hz frequency shift for heliox) that is related to the gas density ($\rho _{g}$) ~\cite{TF_SF6,TF_liquid}. The concentration of carbon dioxide $[CO_{2}]$ is monitored by an infrared absorption detector. A thermistor (G540 10K$\Omega $, Glass-Encapsulated Sensor, EPCOS, Germany) and a pressure sensor (26PCC, honeywell, Canada) have also been used for monitoring the gas temperature and pressure. Oxygen gas concentration $[O_{2}]$ is given by the results from the first few pre-washin breaths, and has been assumed linearly dependent on carbon dioxide concentration, based on the assumption of steady gas exchanging rate during the washout measurement.

\begin{figure}[tbp]
\centering 
\includegraphics[width = \textwidth]{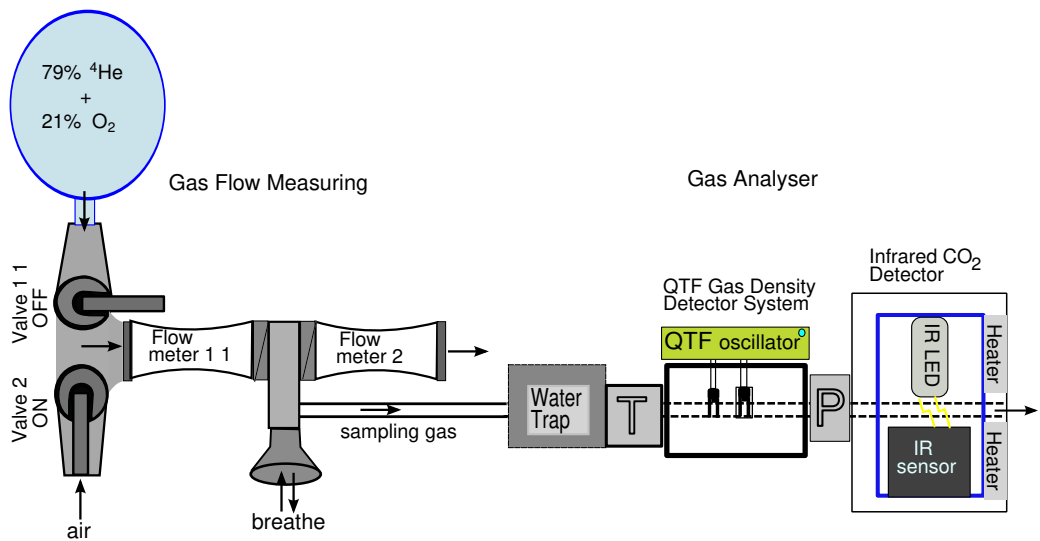}
\caption[MBHW System]{MBHW System. It consists of gas flow
measuring sub-sytem and gas analysing sub-system. Gas flow measuring system consists of two flow meters and two manual vlaves for controlling the inhaling gas (air or heloix). The sampling gas is taken nearby the mouth and then flows through a water trap which is connected with the gas analyser. The gas analyser mainly consists of the QTF gas density detector and a infrared CO$_2$ detector. A thermistor (T) and a pressure sensor (P) are also connected to the sampling tube. }
\label{MBHW System}
\end{figure}

Helium gas concentration $[He]$ can thus be given by subtracting other gas
concentrations with two relations of 
\begin{equation}
[He]=1-[N_{2}]-[H_{2}O]-[O_{2}]-[CO_{2}],
\end{equation}%
and 
\begin{equation}
\rho _{g}=\rho _{He}[He]+\rho _{N_{2}}[N_{2}]+\rho _{H_{2}O}[H_{2}O]+\rho
_{O_{2}}[O_{2}]+\rho _{CO_{2}}[CO_{2}],
\end{equation}%
with $[H_2O]$=0.61, 0$\leq[CO_2]\leq$6, and $[O_2]$is a linear function of $[CO_2]$.

\subsection{Instrumentation}
The working priciples of quartz-tuning-fork (QTF) gas density detector and the carbon dioxide infrared detector are described in this section.

\subsubsection{Quartz Tuning Fork Density Detector}
\begin{figure}[tbp]
\centering 
\includegraphics[]{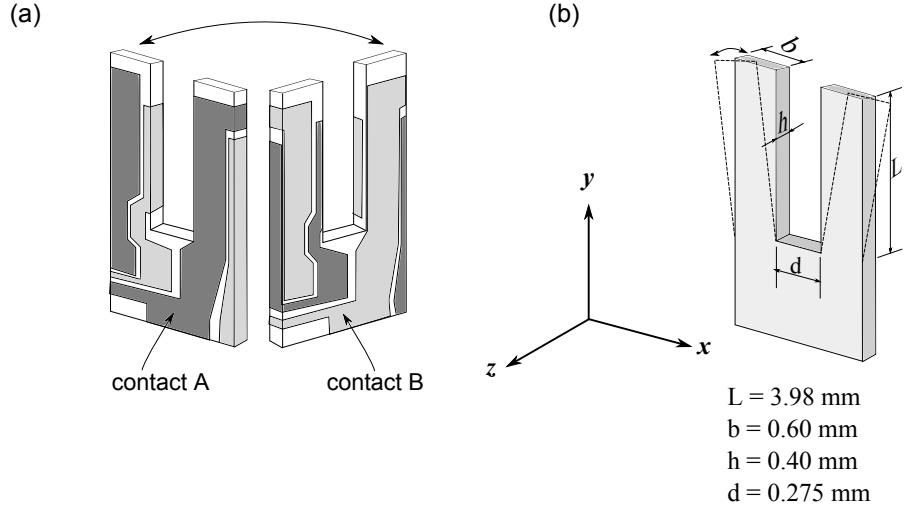} 
\caption[Sketch of Quartz Tuning Fork]{(a) Sketch of the quartz
tuning fork and (b) the main flexural resonating mode of ~32768 Hz.}
\label{Tuning Fork}
\end{figure}

Piezoelectric quartz tuning forks for time-keeping element of digital
watches are used in this study. Because of their high stability and quality factor (10,000 in air), quartz tuning forks have been used in various sensor applications; for example, measuring the thermal properties of liquid helium~\cite{TF_for_helium}, as force sensors for atomic force microscopy~\cite{TF_AFM,AFM_saitoh, AFM_yoshiaki, AFM_hasegawa,AFM_pinto, AFM_kunstmann}, as SF$_6$-gas density sensors~\cite{TF_SF6}, and as mass sensitive sensors in liquid~\cite{TF_liquid}. The mechanical resonance of the tuning fork geometry is coupled electrically to contacts patterned on the surface of the quartz (figure 6) ~\cite{beam_vib}. When a tuning fork is placed in a fluid medium, each tine can be regarded as a damped oscillating cantilever ~\cite{Chu_TF,Sader2006}. The equation of motion can be given by the form of harmonic oscillator for the fundamental mode with a driving force \textit{F} at angular frequency $\omega $, i.e., 
\begin{equation}
Fe^{i\omega t}=m_{e}\frac{\partial ^{2}x}{\partial t^{2}}+m_{e}\gamma \frac{%
\partial x}{\partial t}+kx,
\end{equation}%
where $m_{e}$ is the effective mass of the cantilever, $x$ is the relevant
vibrating displacement, $\gamma $ is the drag force constant caused by the
viscosity of the fluid, and $k$ is the elastic constant of the cantilever.
For a uniform cantilever beam with rectangular cross section, $k$ can also
be expressed in terms of the Young's modulus $E$ by~\cite{Cleveland_TF} 
\begin{equation}
k=E\frac{h^{3}b}{4L}
\end{equation}%
where $h$, $L$, and $b$ are the dimensions of the cantilever beam as shown
in figure 6. The related fundamental resonance frequency $f$ can be
expressed in a general form of 
\begin{eqnarray}
f=\frac{\omega }{2\pi } &=&\frac{1}{2\pi }\sqrt{\frac{k}{m_{e}}} \\
&=&\frac{1}{2\pi }\sqrt{\frac{Eh^{3}b}{4Lm_{e}}} \\
&=&\frac{C_{q}}{2\pi }\sqrt{\frac{E}{\rho _{q}}{\rho _{g}}},
\end{eqnarray}%
where $C_{q}$ is the geometry factor of the cantilever, $\rho _{q}$ and $%
\rho _{g}$ are the density of the cantilever (quartz) and the surrounding
medium (gas), respectively. For small gas density $\rho _{g}$ compared to
the density of quartz $\rho _{q}$, the frequency shift ($\Delta f_{g}$) can
be expressed as 
\begin{eqnarray}
\frac{\Delta f_{g}}{f_{v}}& = & \frac{f_{v}-f_{g}}{f_{v}} \\
 & = & 1-\sqrt{\frac{\rho _{g}}{\rho _{q}+\rho _{g}}} \\
& \cong &  \frac{1}{2}\left( \frac{\rho _{g}}{\rho _{q}}\right) ,
\end{eqnarray}%
where the resonance frequency in vacuum ($f_{v}$) and in gas ($f_{g}$) are
given by 
\begin{eqnarray}
f_{v} &=&f(\rho _{g}=0)=\frac{C_{q}}{2\pi }\sqrt{\frac{E}{\rho _{q}}}, \\
f_{g} &=&f(\rho _{g}>0)=\frac{C_{q}}{2\pi }\sqrt{\frac{E}{\rho _{q}+\rho _{g}%
}}.
\end{eqnarray}%
The resonance frequency in heliox-air mixture ($f_{He}$) can be related to
the resonance frequency in nitrogen ($f_{N_{2}}$)and carbon dioxide ($%
f_{CO_{2}}$) as 
\begin{equation}
\frac{\Delta f_{He}-\Delta f_{N_{2}}}{f_{v}}\cong \frac{1}{2}\left( \frac{%
\rho _{He}-\rho _{N_{2}}}{\rho _{q}}\right) ,
\end{equation}%
\begin{equation}
\frac{\Delta f_{He}-\Delta f_{N_{2}}}{\Delta f_{CO_{2}}-\Delta f_{N_{2}}}%
\cong \left( \frac{\rho _{He}-\rho _{N_{2}}}{\rho _{CO_{2}}-\rho _{N_{2}}}%
\right) ,
\end{equation}%
where $\rho _{He},\rho _{N_{2}},\rho _{CO_{2}}$ are the densities of
heliox-air mixture, nitrogen, and carbon dioxide, respectively. For heliox
(79\% He + 21\% O$_{2}$). the frequency shifts about 4 Hz compared to room
air.

The net effect of the surrounding medium will cause a change in the
effective mass of the cantilever which depends on fluid density and
viscosity. In this work, instead of measuring the frequency in vacuum ($f_v$%
), the resonance frequency in pure nitrogen $f_{N_2}$ and in pure carbon
dioxide $f_{CO_2}$ are measured as the reference frequencies before every
the washout measurement to allow accurate calculation of expiratory gas
density. Since the density of helium is much lower than of nitrogen ($%
(\rho_{He}-\rho_{N_2})/\rho_{N_2}=-86\%$) compared to the viscosity ($%
(\mu_{He}-\mu_{N_2})/\mu_{N_2}=5.6\%$), only the density effect is taken
into account in the frequency shift.
\newline

The quartz tuning forks (Watch Crystal NC26, Fox Electronics, USA) are
commercially supplied in a sealed can within a small amount of helium gas.
In this work, the quartz tuning fork is used as an actuator driven by a Pierce oscillator. Its resonance frequency is measured when it is exposed in the expired gas mixtures with the vacuum can removed.

The QTF gas density detecting system diagram is shown in figure 7.
Two QTFs are mounted in a plastic cell where gas passes through, one is
exposed to the gas and the other one is left with its vacuum can in place.
The one within its vacuum can is for compensating any change due to the
temperature variation and providing a reference frequency signal. Two Pierce oscillator circuits have been built for each QTF and fixed on a circuit board attached to the QTF cell and sealed with two rubber
o-rings. A commercial frequency detector (easyPLL plus 3.0 Detector,
NanoSurf, Switzerland) with an inner phase lock loop (PLL) has been used for detecting the
frequency shift from the QTF. 
\begin{figure}[tbp]
\centering  
\includegraphics[]{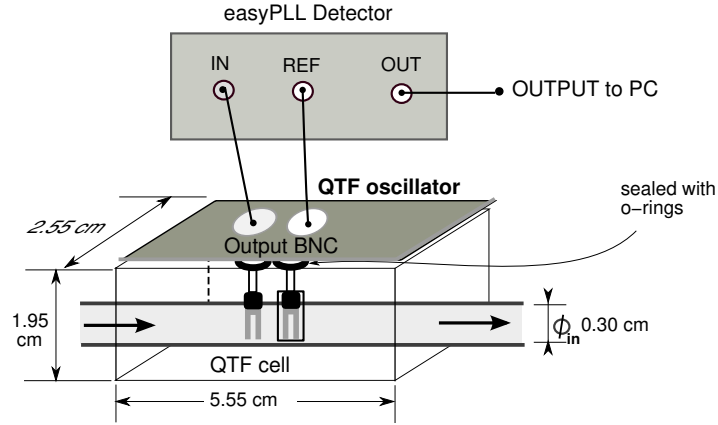}
\caption[QTF Gas Density Detecting System]{QTF gas density
detecting diagram.}
\label{QTF Gas Density Detecting System}
\end{figure}

\subsubsection{Infrared Carbon Dioxide Analyser}
The expired gas from our breath contains around 5\% of CO$_{2}$. 4.2-$\mu$m infra-red light has been used for detecting the concentration of carbon dioxide which is sensitive to the CO$_{2}$ but not to the water vapour~\cite{Infrared_abs,IR_CO2}. The infrared absorption by carbon dioxide follows the Beer-Lambert's law in optics which describes the fact that the absorption of light by a material is dependent on the concentraion and thickness of the material, $i.e.$,~\cite{Infrared_abs} 
\begin{equation}
I_{out}=I_{0}e^{-\kappa l}
\end{equation}%
where $I_{0}$ is the initial light intensity while $I_{out}$ is the output
light intensity, $\kappa $ is the absorption coefficient of the material,
and $l$ is the thickness of the material travelled by the light. For
multiple materials, the absorption coefficient $\kappa $ is a linear
combination of all compositions, $i.e.$ 
\begin{equation}
\kappa =c_{1}\epsilon _{1}+c_{2}\epsilon _{2}+c_{3}\epsilon _{3}+\cdots
=\sum_{i}c_{i}\epsilon _{i}
\end{equation}%
where $c_{i}$ and $\epsilon _{i}$ are the concentrations and extinction
coefficients of the materials, respectively.

For a small change of gas concentration, the change of the exponential term ($e^{-\kappa l}$) can be expressed in a linear form ($1-\kappa l$) and thus the output light intensity can be linearly related to the gas concentration. The photoconductive PbSe film converts the optical energy into the electrical current and thus the concentration of carbon dioxide can be linearly related to the output voltage~\cite{US6218665,US4578762}.

\begin{figure}[tbp]
\centering 
\includegraphics[]{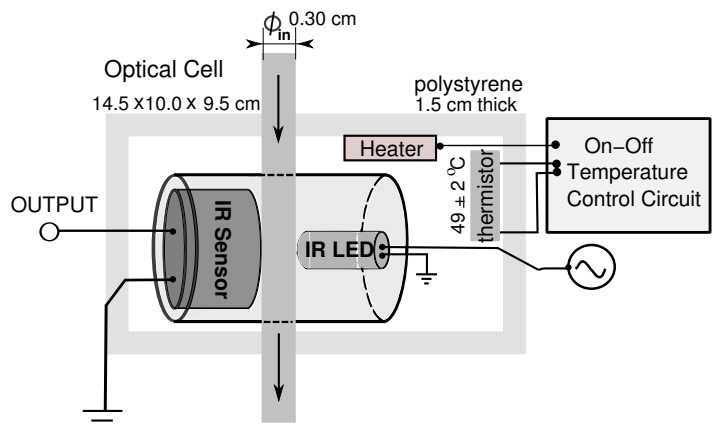}
\caption[Optical System - Infrared CO$_{2}$ Detector]{
The infrared CO$_{2}$ detecting cell with a temperature-control circuit.}
\label{Infrared CO$_2$ Detecting System Diagram}
\end{figure}
As shown in figure 8, the CO$_2$ detector contains a 4.2 $\mu$m infrared LED (LED42SC, Scitec,UK) and the PbSe photoconductive detector (P9696 series, Hamamatsu, Japan) insulated from the surroundings by 1.5 cm-thickness of expanded polystyrene. The optical cell is temperature controlled to 49 $^o$C to improve stability and to prevent the condensation of the water inside the cell.

\subsection{Calibration}

\subsubsection{QTF in Heliox-Air Mixture}

\begin{figure}[tbp]
\centering 
\includegraphics[width = 10cm]{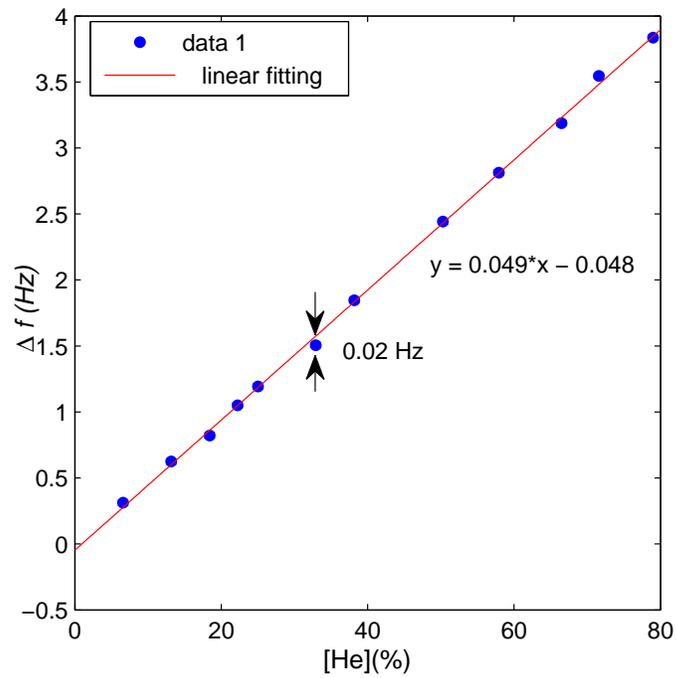}
\caption[Helium Concentration versus Resonance Frequency Shifts]{The calibration of resonance frequency shifts $\Delta f$ in different concentrations of heliox-air mixture. It shows a linear relation
with a corresponding 0.02-Hz digital noise.}
\label{Calibration for Heliox-Air Mixture}
\end{figure}

The relationship between the resonance frequency shifts and the helium gas
concentration in air has been found by monitoring varying the known gas
mixture composition. A linear relation is shown in figure 9 with the signal noise which corresponds to 0.02 Hz.

\subsubsection{QTF and CO$_2$ Sensor in N$_2$-CO$_2$ Mixtures}
During the helium washout measurements, the flowing gas mixture passes
through the cold water trap, the QTF cell, the thermistor and pressure
sensor, and the infrared CO$_{2}$ cell. To get the density of the
gas mixture in the flowing gas, a calibration has been performed by putting
the 100\% flowing nitrogen and 5.6\% carbon dioxide gas through the gas
analysing system before every washout measurement. Thus, the resonance
frequency shifts $\Delta f_{N_{2}},\Delta f_{CO_{2}}$, and $\Delta f_{g}$
can be found. The corresponding temperature, pressure and signal from the
carbon dioxide detector were also monitored at the same time.
Figure 10 shows the singals from the QTF, infrared CO$_2$ detector, pressure sensor, and thermistor with 100\% N$_{2}$, 5.6\% CO$_{2}$, and 100\% CO$_{2}$ gas flowing through including values of the corresponding noise. 

\begin{figure}[tbp]
\centering 
\includegraphics[width=\textwidth]{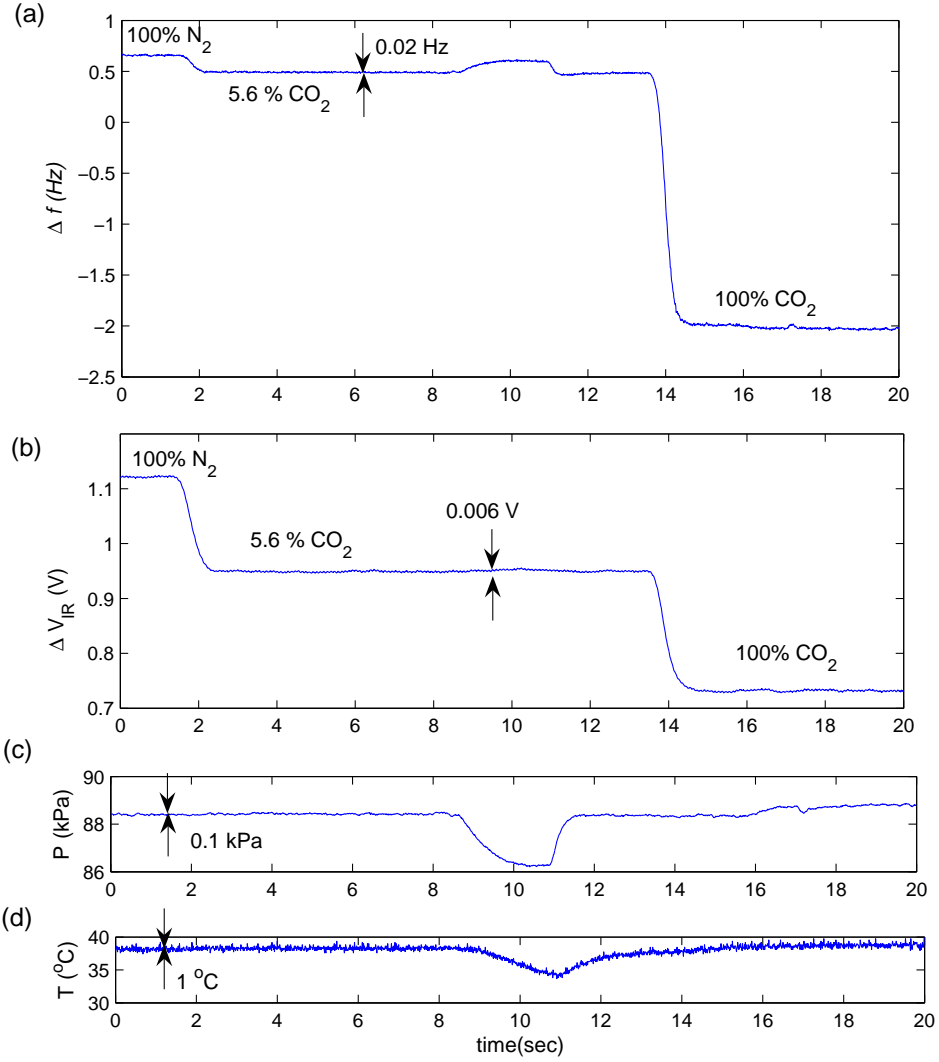}
\caption[Lower Concentration of CO$_{2}$ Calibration Curve]{
The output signals and the corresponding signal noise from (a) QTF fequency shift,  (b) infrared carbon dioxide detector, (c) pressure sensor, and (d) the thermistor when 100\% nitrogen, 5.6\% carbon dioxide, and 100\% carbon dioxide flowing through.}
\label{Signal Noise}
\end{figure}

\subsubsection{Oxygen Concentration in Pre-washout Measurement}

Before MBHW washout measurements (before breathing in heliox), a few normal
tidal breaths are taken to measure the gas exchanging rate and oxygen
concentration. The concentration of oxygen can be given by monitoring the
gas density from the gas mitxture (N$_{2}$, O$_{2}$, H$_{2}$O, CO$_{2}$) and
the carbon dioxide concentration with 0.61\% of water vapour. Two linear
phases have been found by plotting the CO$_{2}$ concentration versus O$_{2}$
concentration in each single expiration during pre-washout measurement shown
in figure 11 with the corresponding signal noise. The first
short phase is related to the system dead space where there is no gas
exchange. The second linear phase has been defined as the gas exchanging
rate in this study, $i.e.$, the consumption of oxygen versus the production
of carbon dioxide. The expired carbon dioxide concentration profile varies
little from breath to breath during the washout measurement which implies
that the existence of the helium does not affect the gas exchange.

\begin{figure}[tbp]
\centering 
\includegraphics[width = \textwidth]{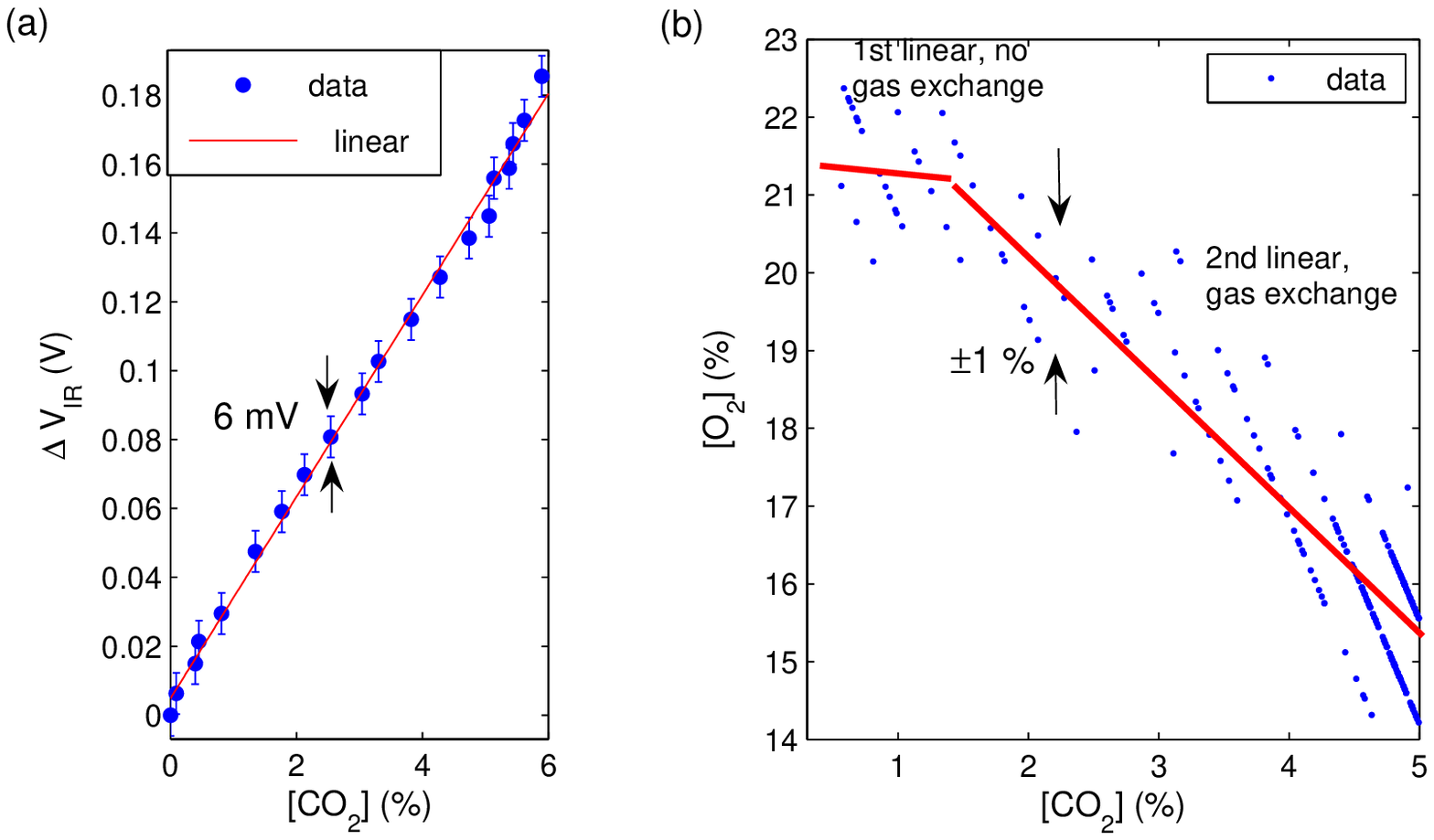} 
\caption[Systematic Error]{Systematic error caused by the
sensor signal noise from two main calibrations. (a) is the concentration of
carbon dioxide versus output infrared signal with 6mV noise. (b) is the oxygen concentration calibrated with both QTF and optical sensors with 2\% [O$_2$] error from a normal breath. }
\label{Systematic Error}
\end{figure}

\subsubsection{Signal Noise}
In this work, the infrared CO$_2$ detector was calibrated by passing through different concentrations of CO$_2$-N$_2$ gas mixture ([CO$_2$] from 0 to 5.6\%) with 6mV digital noise. The oxygen concentration is given by calibrating the pre-washout signal with infrared sensor and QTF. The highest signal noise from two sensors results in 2.0\% [O$_2$] of error in room air which contains around 21.0\% of oxygen. 

\section{Results}

\subsection{Lung Clearance}

The lung clearance results from a mild asthmatic and a normal subject are
shown in figure 12 with two fitted exponential curves representing
compartments with better and poorer ventilation. The result from the normal
subject shows a nearly ideal trace with one ventilation rate. However, for
the mild asthmatic subject, two compartments with very different ventilation
rates that can easily be differentiated and occupy 33\% and 67\% of the lungs.

\begin{figure}[tbp]
\centering 
\includegraphics[width = \textwidth]{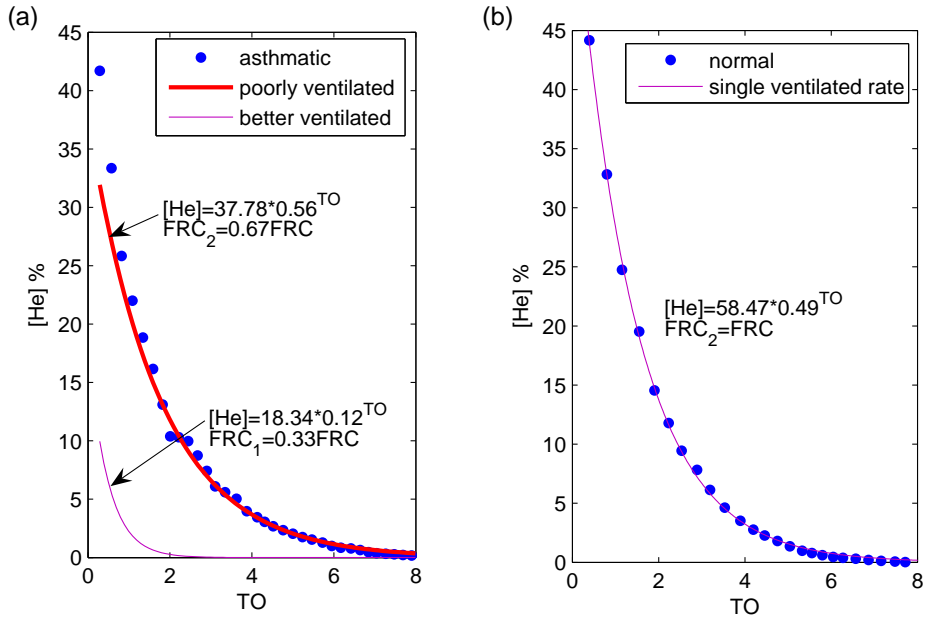}
\caption[Lung Clearance]{(a) Lung clearance result from a
mild asthmatic female with two fitted exponential curves. The lung is
occupied by 67\% of a better-ventilated compartment ($FRC_1$, thick curve)
and 33\% of less-ventilated part ($FRC_2$, thin curve). (b) Lung
clearance result from a normal female an exponential curve. The lung
is occupied by only one compartment ($FRC_1$, thin curve) from the results of fitting with two exponential curves. }
\label{Lung Clearance Curve - Asthmatic Female}
\end{figure}

The peak expiratory flow (PEF) has been measured as a clinical reference with a Wright metre (Mini-Wright, Standard Range, EN13826, Clement Clarke
International, UK) on all volunteers with the predicted values from the
metre data sheet\footnote{This can be found on the website \textit{http://www.peakflow.com/top\_nav/normal\_values/index.html}}. The PEF measurement has been regarded as a sensitive technique for identifying airflow obstruction ~\cite{PEF}. From our MBHW results, the measured lower PEF values are not necessarily related to the lung obstruction but somehow reflected in a larger proportion of poorly ventilation compartment ($V_{poor}/FRC$) with very different ventilation rates. For example, the PEF values for two male subjects (No.1 and 3) are much smaller than the predicted values, and both of them have very different ventilation rates. However, the measured PEF value for the asthmatic subject (No.11) is higher than the predicted value which can be a result of regular exercise habbits, while both of the compartments and ventilation rates are apparently different.

\begin{sidewaystable}[] 
\caption{\textbf{MBHW Results - Lung Clearance}}
{\centering{ 
\linespread{1.8}
{\small{
\begin{tabular}{l|c|cc|c|ccc|cc|c|c}
\hline
No. & Gender, & Height & Weight & $PEF^m/PEF^p$ & $Q_r$ & FRC & V$_T$ & $V
_{better}/FRC$ & $V_{poor}/FRC$ & $R_{better}/R_{poor}$ & $LCI_{He}$\\ 
 & Age(yrs) & (cm) & (kg) & (\%) &  & (L) & (L) &  &  & & \\ \hline
1& M, 25 & 173 & 100 & \textbf{79.77} & 1.50$\pm$0.13 & 4.16 & 1.02 & 0.41 & \textbf{0.59} & \textbf{1.8421} & 5.64\\ 
2& M, 24 & 178 & 66 & 95.04 & 1.65$\pm$0.06 & 2.76 & 0.91 & 0.93 & 0.07 & 2.9174 & 5.15\\ 
3& M, 25 & 195 & 75 & \textbf{66.72} & 1.30$\pm$0.11 & 5.77 & 2.09 & 0.24 & \textbf{0.75}& \textbf{2.2336} & 7.70\\ 
4*& M$_{ah}$, 27 & 175 & 65 & 106.80 & 1.64$\pm$0.09 & 3.61 & 0.92 & 1 & - & 1 & 4.09\\ 
5*& M$_{ch}$, 29 & 185 & 78 & 91.81 & 2.03$\pm$0.08 & 3.58 & 1.49 & 0.84 & 0.16 & 1.6648 & 6.30\\ 
6$^+$& M$_a$, 29 & 165 & 59 & \textbf{75.63} & 1.97$\pm$0.11 & 2.64 & 0.56 & 0.49 & \textbf{0.51} & \textbf{11.0674} & 6.95\\ 
7$^+$& M$_a$, 47 & 175 & 76 & 91.20 & 1.74$\pm$0.09 & 3.71 & 1.52 & 0.68 & 0.32 & 1.5007 & 6.69\\ 
8**& M$_s$, 27 & 175 & 74 & 129.03 & 1.52$\pm$0.07 & 3.40 & 0.88 & 0.99 & 0.01 & 1.5470 & 5.04\\ 
9& F, 26 & 166 & 50 & 95.24 & 2.33$\pm$0.08 & 3.01 & 0.89 & 1 & - & 1 & 4.67\\ 
10& F, 26 & 155 & 57 & 99.30 & 1.90$\pm$0.11 & 1.87 & 0.69 & 0.98 & 0.02 & 1.4131 & 5.93\\ 
11$^+$& F$_a$, 24 & 155 & 54 & \textbf{107.14} & 2.39$\pm$0.08 & 2.03 & 0.49 & 0.33 & \textbf{0.67} & \textbf{4.6002} & 5.88\\ 
12**& F$_s$, 35 & 163 & 53 & 98.88 & 1.60$\pm$0.07 & 1.86 & 0.73 & 0.84 & 0.16 & 1.6829 & 6.85\\ \hline
\end{tabular}
}}}}\\
{Lung Clearance Results from four male (one with asthma history M$_{ah}$, No.4*, and one with COPD history, M$_{ch}$, No.5*) and two female normal volunteers, two male mild asthmatics (M$_a$, No.6$^+$ and 7$^+$) and a female mild asthmatics (F$_a$, No.11$^+$), and a female (F$_s$, No.12**) and male (M$_s$, No.8**) smoker. Two normal male subjects (No.1 and 3) having smaller ratio of the measured PEF values (PEF$^m$) over the predicted values (PEF$^p$) are found to have larger capacities of poorly ventilated volumes $\frac{V_{poor}}{FRC}$. $Q_r$ is the gas exchanging rate defined in section 2.1.}
\label{tab:Lung Clearance Results}
\end{sidewaystable}

\subsection{Normalised Phase III Slopes}

Figure 13 shows the results from two mild asthmatics and two normal subjects in which the normalised phase III slopes of helium during (a) helium washin and (b) helium washout process are plotted versus the turnover. In (a), the wash-in results from a mild asthmatic female shows both of $S_{cond}$ and $S_{acin}$ values are higher than the normal ones which indicates the higher ventilation inhomogeneities in both conductive and acinar airways. In (b), the washout results from a mild asthmatic male shows a higher $S_{cond}$ value than the normal one but similar $S_{acin}$ values which indicates the asthmatic male has no apparent higher ventilation inhomogeneity in the acinar airways than the normal volunteer. 

\begin{figure}[tbp]
\centering 
\includegraphics[width = \textwidth]{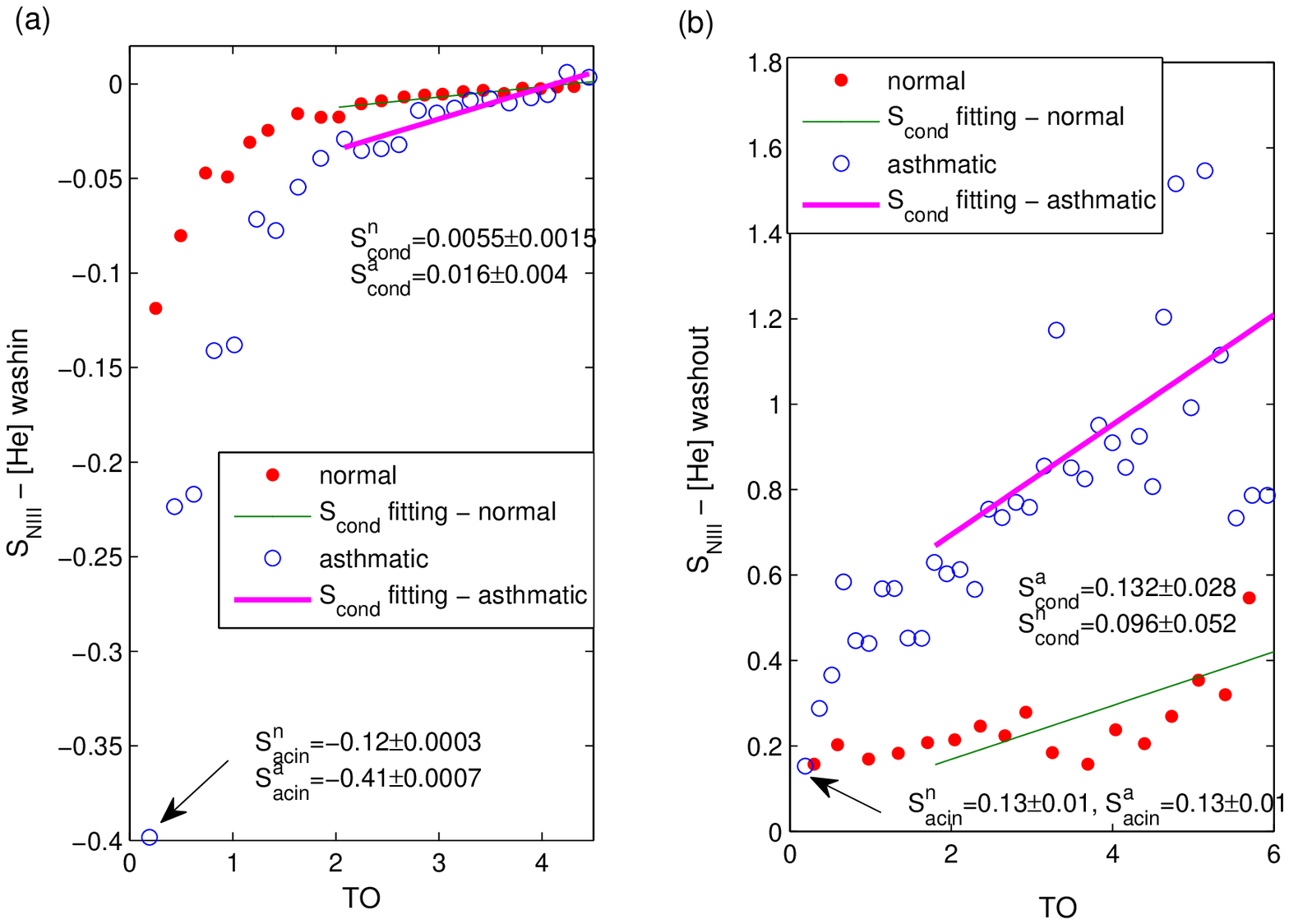}
\caption[$S_{NIII,out}$ - Asthma Male]{(a) The normalised
phase III slopes of helium during helium wash-in process from a mild
asthmatic and a normal female volunteer. The results from the asthmatics show higher absolute values of $S_{cond}$ ($S^a_{cond}=$0.016$\pm$0.004) and $S_{acin}$ ($S^a_{acin}=0.41\pm0.0007$) compared to the normal ones ($S^n_{cond}$=0.0055$\pm$0.0015, $|S^n_{acin}|=0.12\pm0.0003$). The arrow indicates the normalised phase III slope of the first breath from the asthmatics is much steeper compared to the normal volunteer which results in a higher absolute value of $S_{acin}$ and implies higher ventilation inhomogeneity in the acinar airway. (b) The normalised phase III slopes of nitrogen during helium washout process from a mild asthmatic and a normal male. The results from the mild asthmatic show a higher $S_{cond}$ value but similar $S_{acin}$ value compared to the normal ones.}
\label{Washout Phase III Slopes - Asthmatic}
\end{figure}

\begin{table}[tbp]
\centering
\linespread{1.8} {\footnotesize \ {\textbf{MBHW results from all volunteers}}
\begin{tabular}{l|c|c|cc|cc}
\hline
No.   & Gender,      & LCI   & $S^{He,in}_{acin}$   & $S^{He,out}_{acin}$  & $S^{He,in}_{cond}$    & $S^{He,out}_{cond}$  \\ 
      & Age(yrs)     &       & *10$^{-3}$(L$^{-1}$) & *10$^{-3}$(L$^{-1}$) & *10$^{-3}$(L$^{-1}$)  & *10$^{-3}$(L$^{-1}$) \\ \hline
1     & M, 25        &  5.65 & -154.35$\pm$0.71     & 131.61$\pm$10.70     &  2.20$\pm$0.85        & 70.98$\pm$13.8 			\\ 
2     & M, 24        &  5.15 & -195.30$\pm$0.22     & 46.42$\pm$4.20       &  5.75$\pm$0.52        & 56.04$\pm$7.58 			\\ 
3     & M, 25        &  7.70 & -217.30$\pm$0.65     & 33.80$\pm$1.60       &  6.64$\pm$1.66       	& 23.09$\pm$2.57       \\ 
4*    & M$_{ah}$, 27 &  4.09 & -277.79$\pm$2.00     & 139$\pm$11.10        &  9.75$\pm$5.95       	& 40.57$\pm$20.91      \\ 
5*    & M$_{ch}$, 29 &  6.29 & -115.40$\pm$2.24     & 82.30$\pm$2.50       &  2.75$\pm$0.39        & 16.24$\pm$2.87       \\ 
6$^+$ & M$_a$, 29    &  6.95 & -196.43$\pm$0.61     & 135.00$\pm$0.98      &  7.39$\pm$1.91        & 95.91$\pm$5.21       \\ 
7$^+$ & M$_a$, 47    &  6.68 & -240.99$\pm$0.53     & 239.10$\pm$2.10      &  6.01$\pm$0.99        & 26.79$\pm$3.10       \\ 
8**   & M$_s$, 27    &  5.04 & -242.20$\pm$0.24     & 181.40$\pm$5.70      &  3.88$\pm$0.75       	& 45.89$\pm$12.68     \\ 
9     & F, 26        &  4.67 & -120.10$\pm$0.38     & 24.51$\pm$2.09       &  5.49$\pm$0.75       	& 74.04$\pm$26.65     \\ 
10    & F, 26        &  5.93 & -118.80$\pm$0.51     & 128.10$\pm$36.10     &  3.59$\pm$0.55       	& 75.54$\pm$22.52     \\ 
11$^+$& F$_a$, 24    &  5.88 & -401.60$\pm$0.78     & 228.30$\pm$25.30     &  16.45$\pm$2.05      	& 109.16$\pm$43.52    \\ 
12**  & F$_s$, 35    &  6.85 & -377.40$\pm$0.95     & 376.30$\pm$12.3      &  14.06$\pm$2.59      	& 30.29$\pm$21.55      \\ \hline
\end{tabular}
}
\caption[MBHW Resutls - Two Indices and LCI]{\linespread{1.2}{\protect\small {MBHW Results from all volunteers. $S_{cond}$ values during helium wash-in and washouts. The lung clearance index from helium washout results are also listed.}}}
\label{tab:MBHW Resutls}
\end{table}

The $S_{cond}$ and $S_{acin}$ values from the all the volunteers have been
listed in table 2. The results from four groups (normal, normal with
asthma and COPD history, smoker, mild asthmatic) are compared and plotted in
figure 14 and 15. The mild asthmatic group shows higher $S_{cond}$ and $S_{acin}$ values in
wash-in but not much difference in washout measurements compared to the rest
of the groups. For mild asthmatics, subject No.6 has higher $S_{cond}$ values while subject No.7 has higher $S_{acin}$ values, and No.11 has higher values in both $S_{cond}$ and $S_{acin}$ compared to the normal group.
For the two smokers, the $S_{acin}$ values are higher compared to the normal-subject group which reflects with ventilation inhomogeneity in the acinar airways.


\begin{figure}[tbp]
\centering 
\includegraphics[width = \textwidth]{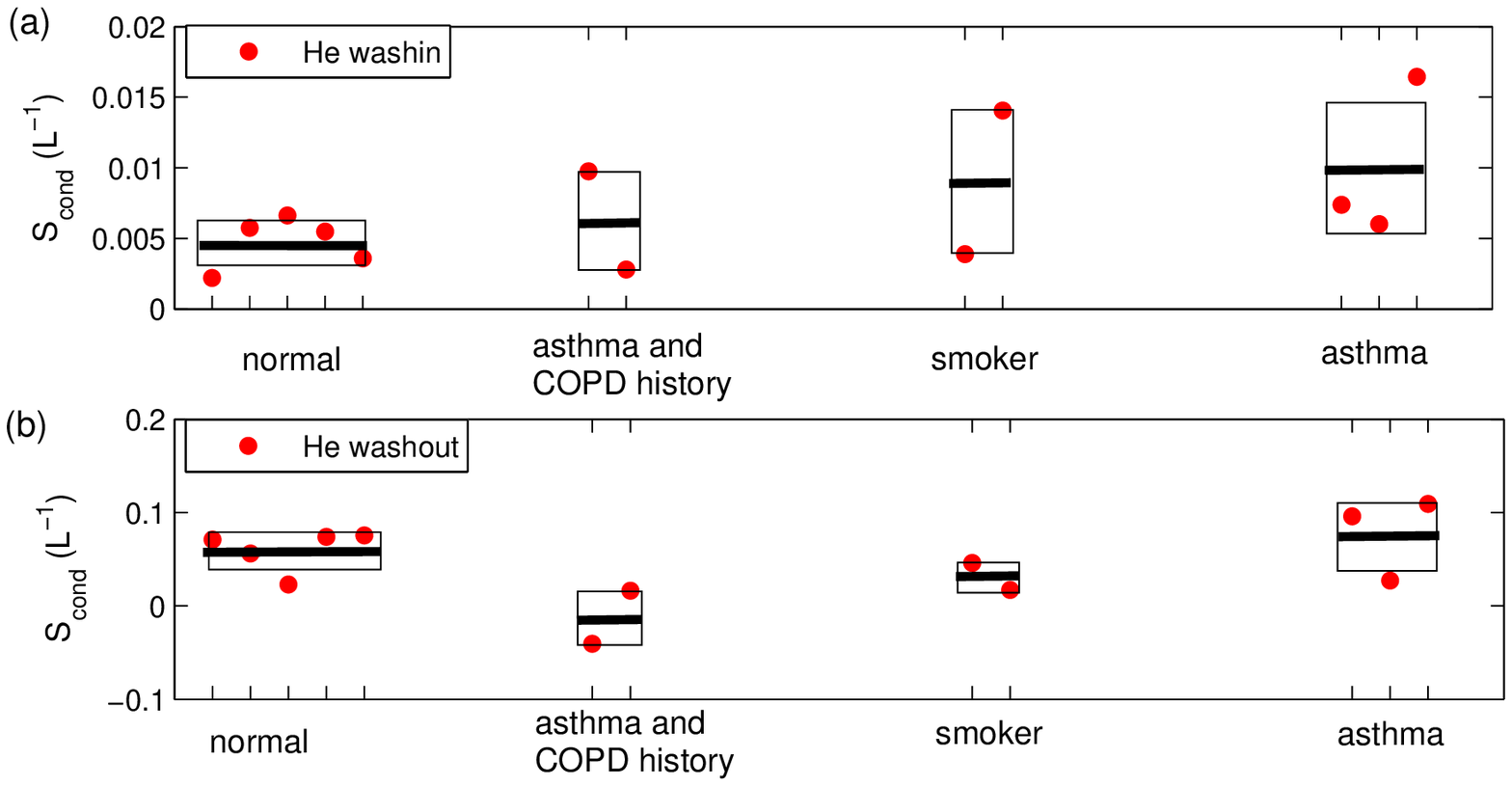}
\caption[MBHW results from all volunteers]{MBHW results of $S_{cond}$ values from all volunteers into four groups where the errors have taken the standard deviation for each group with the mean values. (a) is from the helium wash-in and (b) is from the helium washout results. Mild asthmatic group has shown a slightly higher mean value than normal group.}
\label{MBHW Results from All Volunteers - $S_{cond}$}
\end{figure}

\begin{figure}[tbp]
\centering 
\includegraphics[width = \textwidth]{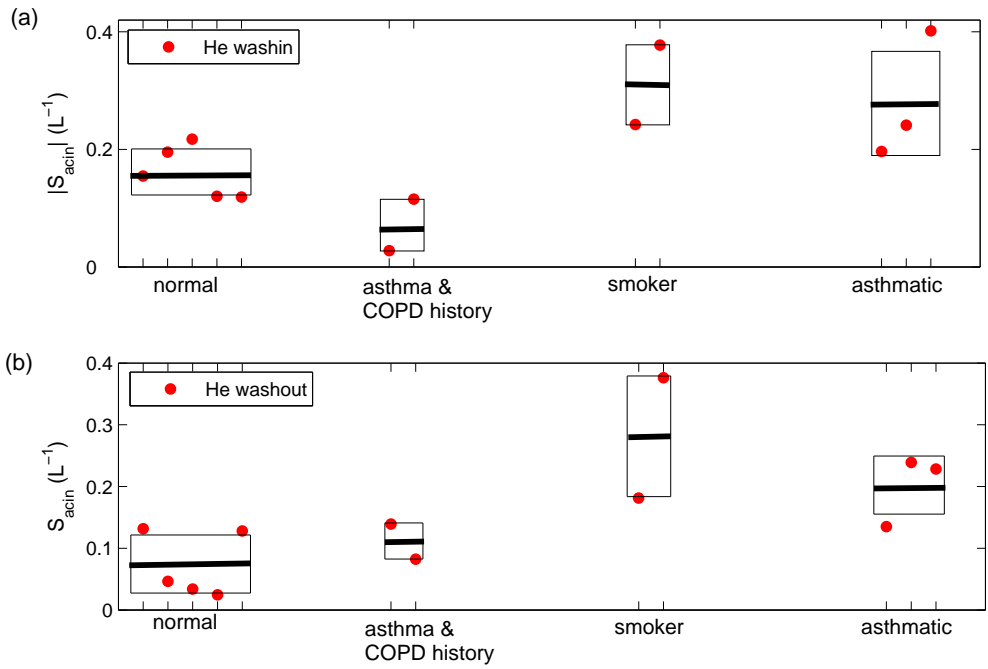}
\caption[MBHW results from all volunteers]{MBHW results of $%
S_{acin}$ values from all volunteers into four groups where the errors have taken the standard deviation for each group with the mean values. (a) is from the helium wash-in results and (b) is from the helium washout results. Results show that smokers have higher $S_{acin}$ values in both wash-in and washouts.}
\label{MBHW Results from All Volunteers - $S_{acin}$}
\end{figure}

\subsection{Smokers}
 
\begin{figure}[tbp]
\centering 
\includegraphics[width = \textwidth]{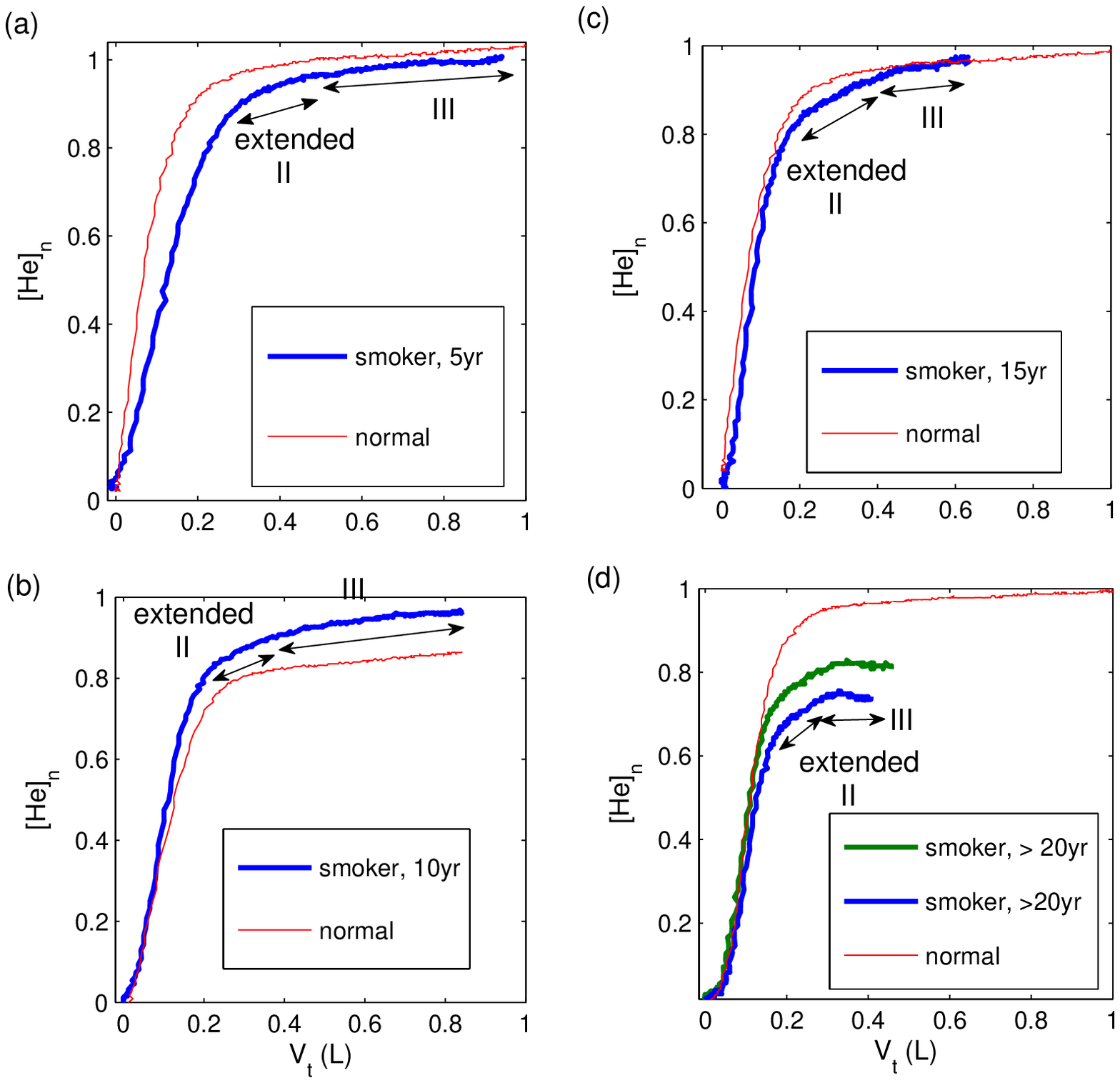}
\caption[Single Curve from Smokers]{The single breath curves from MBHW measurements with four smoker volunteers having different length period of smoking history compared to the normal subjects. For smokers there is a more extended phase II which causes the phase III linear region less clearly defined. For smokers with longer period of smoking history ($>$10 years), the feature becomes more evident which implies a higher uneven ventilation between conductive and acinar airways.}
\label{Single Curve from Smokers}
\end{figure}

An extended phase II and delayed phase III have been observed in most of the single breaths from four smokers as shown in figure 16. For smokers with longer smoking history ($>$10 years), this feature becomes more apparent which implies higher uneven ventilation between conductive and acinar airways. The curved transition to phase III, which is little documented in nitrogen washout measurements, results in a difficulty of defining phase III but may imply an early stage of phase III steepening which has been observed in COPD patients. Also, small and unsteady tidal volumes have also been observed from four smokers compared to the normal volunteers with similar FRC. The helium washout results from two of the smokers have to be excluded because of their constant coughing or short of breath during the measurements (10- and 20-year smoking history). The washout results from the other two smokers (5- and 15-year) have included the curved transition into phase III.

\section{Conclusion}

The MBHW system containing a QTF gas density sensor and a infrared CO$_{2}$
detector has been developed for MBW lung function study. The system gives
accurate values of gas concentrations and can be used to extract the $LCI$
as well as $S_{cond}$ and $S_{acin}.$The MBHW results from 12 volunteers
show a difference between normal and asthmatic (or smokers) subjects.
Subjects with lower PEF values (compared to the predicted values) are found
to have smaller portion of better-ventilated lung capacity, which is not
necessarily related to lung disease. 

Higher $S_{cond}$ values or $S_{acin}$ values have been found on mild asthmatics compared to normal
subjects which reflect the inhomogeneis ventilations in mildly diseased
lungs. The higher $S_{acin}$ values have also been found on the smoker with 15-year smoking history. 


The results from the wash-in process show a more apparent difference
in $S_{cond}$ and $S_{acin}$ between the four groups indicating that it may
be a more sensitive manuevour compared to the washout process.

An extended phase II and delayed phase III have been observed in single breath curves from four smokers with different length of smoking history. For smokers with longer smoking history ($>$10 years), this feature becomes more apparent which implies higher uneven ventilation between conductive and acinar airways and may indicate the early stage of phase III steepening observed from COPD patients.

\appendix

\section{Lung Clearance}
\label{Lung Clearance}

The washout lung clearance curve is taken by plotting the mean concentration
of gases (helium or nitrogen) as a function of turnover $TO$. The mean
concentration of gases ($c_n$ from $nth$ breath) have been taken as the
expired amount of gas ($\int^{V_T}_{0}c_ndV$) divided by the expired tidal
volume ($V_T$) from each breath. The turnover $TO$ is the accumulative
expired tidal volume ($\sum V_T$) divided by the functional residual
capacity ($FRC$). In the helium or nitrogen washouts, $FRC$ has been simply
derived from 
\begin{eqnarray}
c_1 &=& \frac{FRC\cdot c_0}{FRC+V_{T,1}} , \\
c_2 &=& \frac{FRC\cdot c_1}{FRC+V_{T,2}} , \\
\vdots \\
c_n &=& \frac{FRC\cdot c_{n-1}}{FRC+V_{T,n}} , \\
FRC &=& \frac{\sum^{n}_{i=1} c_i\cdot V_{T,i}}{c_0-c_n},
\end{eqnarray}
where $c_{n=0,1,2,\cdots n-1}$ and $V_{T,n (n=0,1,2,\cdots n-1)}$ are the
mean concentration of gas and tidal volumes from the initial gas in the
lungs, the first, second, and $(n-1)th$ expiration, respectively. This
equation is based on the assumption of well-mixed gases in the lungs. Since
the FRC has been defined as the lung volume at the end of expiration, it is
dependent on the expired tidal volumes and is slightly different breath by
breath. In this study, only gas concentrations higher than 0.1\% have been
taken into acount for the $FRC$ calculation.

On the basis of the two-compartment model, human lungs can be abstractly
regarded as two parallel compartments with different ventilations~\cite%
{Bouhuys1956}. In this section, the washout results from the
helium washout process will be discussed. The mean concentration of helium
from each breath is plotted as a dual-exponential function (the summation of
two exponential functions) of turnover (cumulative expired tidal volumes
divided by FRC). It can be expressed as 
\begin{eqnarray}
FRC &=& FRC_1+FRC_2, \\
TO_n &=& \frac{\sum^{n}_{1} V_T}{FRC}, \\
V_T &=& V_1+V_2, \\
c(1+2,n) &=& \frac{FRC_1}{FRC_1+FRC_2}c_{1,n}+\frac{FRC_2}{FRC_1+FRC_2}c_{2,n}, \\
c_{1,n} &=& c_{a0}\left(\frac{FRC_1}{FRC_1+V_1}\right)^{TO}, \\
c_{2,n} &=& c_{a0}\left(\frac{FRC_2}{FRC_2+V_2}\right)^{TO},
\end{eqnarray}
where $FRC_1$ and $FRC_2$ are the residual capacities of two compartments $1$
and $2$, $c_{1,n}$ and $c_{2,n}$ are the mean helium concentration from the $%
n$th breath, $V_1$ and $V_2$ are the expired tidal volumes of two
compartments. If compartment $1$ represents the better-ventilated
compartment compared to the compartment $2$, then $1$ has a higher washout
rate, e.g., $\frac{FRC_1}{FRC_1+V_1} > \frac{FRC_1}{FRC_1+V_1}$. The ratio
of $\frac{FRC_1}{FRC}$ represents the proportion of the better-ventilated
part of the lung.

In this work, breaths from the first expiration to the end ($> 0.1\%$) of
helium gas concentration during helium washout have been taken into account
for this two-exponential curve fittings. The intial gas concentraion $c_{a0}$
is thus the mean gas concentration from the first breath. If the initial gas
concentration was taken as the gas concentration in the lungs before the
washout measurement, the lung clearance washout curve was unable to be
fitted well with a sum of two exponential curves. This is likely to result
from the imperfect gas mixing such that the mean concentration of the
expired gas is smaller than the mean concentration of gas from the whole
lungs. $FRC_1$, $FRC_2$ are given by Eq. A.9 and $V_1$, $V_2$ are given by Eq.
A.10 and A.11. 

For more complex models with 50 compartments can be found in the studies by Lewis et al. ~\cite{Lewis1978c, Lewis1978e}. A model of tidal emptying pattern has also been developed in a few modern studies by Whiteley et al. with the consideration of gas re-breathing from the dead space ~\cite{Whiteley1999,Hahn1998,Farmery2000}. The mathematical model based on the one- and 50-compartment model are described and compared with the results from the inert gas elimination technique. Unlike the MBW technique, the tracer gas is inserted into the blood and the concentration of the tracer gas is sampled by taking the blood sample. From the inert gas elimination results, the breathing pattern is more likely tidal emptying instead of continuous which means the gas re-breathing from the dead space is unable to be ignored.




\bibliographystyle{model1a-num-names}
\bibliography{MEP2011bib}







\end{document}